\documentclass[12pt]{article}
\usepackage{a4wide}
\usepackage{authblk}
\usepackage{hyperref}
\usepackage[top=25truemm,bottom=25truemm,left=25truemm,right=25truemm]{geometry}
\usepackage{setspace}
\usepackage{amsmath,amssymb,mathrsfs,amsbsy,latexsym,amsfonts,amsthm}
\usepackage{physics}
\usepackage{fancyhdr}
\usepackage[Symbol]{upgreek}

\usepackage{graphicx}           

\usepackage{color}
\usepackage{slashed}
\usepackage{tensor}

\newcommand{\al}[1]{\begin{align}#1\end{align}}

\newcommand{\paren}[1]{\left(#1\right)}

\newcommand{\sqbr}[1]{\left[#1\right]}
\newcommand{\br}[1]{\left\{#1\right\}}

\newcommand{\nn}{\nonumber\\}

\newcommand{\p}{\partial}
\newcommand{\Slash}[1]{{\ooalign{\hfil/\hfil\crcr$#1$}}} 
\newcommand{\outst}{\tensor[_0]{\bra{\text{out}}}{}}
\newcommand{\inst}{\ket{\text{in}}_0}

\usepackage{epsf}

\makeatletter

\@addtoreset{equation}{section}
\makeatother

\title{\bf 
Soft pion theorem, asymptotic symmetry \\and new memory effect
\vspace{5mm}}

\author[a,b]{
Yuta~Hamada\thanks{\tt yhamada@wisc.edu}
}
\author[c]{
Sotaro~Sugishita\thanks{\tt sugishita@het.phys.sci.osaka-u.ac.jp}
\vspace{5mm}
}
\affil[a]{\it\normalsize Department of Physics, University of Wisconsin-Madison, Madison, WI 53706, USA\\
\vspace{1mm}}
\affil[b]{\it\normalsize KEK Theory Center, IPNS, KEK, Tsukuba, Ibaraki 305-0801, Japan\\
\vspace{1mm}}
\affil[c]{\it\normalsize Department of Physics, Osaka University, Toyonaka, Osaka, 560-0043, Japan}
\date{}
\begin{document}
\maketitle
\thispagestyle{fancy}
\renewcommand{\headrulewidth}{0pt}

\begin{abstract}
	It is known that soft photon and graviton theorems can be regarded as the Ward-Takahashi identities of asymptotic symmetries. In this paper, we consider soft theorem for pions, \textit{i.e.}, Nambu-Goldstone bosons associated with a spontaneously broken axial symmetry. The soft pion theorem is written as the Ward-Takahashi identities of the $S$-matrix under asymptotic transformations. We investigate the asymptotic dynamics, and find that the conservation of charges generating the asymptotic transformations can be interpreted as a pion memory effect. 
\end{abstract}

\newpage

\setcounter{tocdepth}{2}
\tableofcontents

\newpage
\section{Introduction}

Recently it has been discussed in \cite{Strominger:2013lka, Strominger:2013jfa, He:2014laa, Kapec:2014opa, He:2014cra,Campiglia:2015qka, Kapec:2015ena,Campiglia:2015kxa,Campiglia:2016hvg,Campiglia:2016jdj,Conde:2016csj,Campiglia:2016efb,Conde:2016rom} (see \cite{Strominger:2017zoo} for a review) that  asymptotic symmetries for QED and Quantum Gravity (QG) in four-dimensional flat spacetime are related to soft photon and graviton theorems \cite{Low:1954kd, GellMann:1954kc, Weinberg:1965nx}. 
The asymptotic symmetries are large gauge transformations for QED and supertranslations in the BMS transformation \cite{Bondi:1962px, Sachs:1962wk} for QG. 
The symmetries are spontaneously broken and soft photons and gravitons can be regarded as the associated Nambu-Goldstone (NG) bosons.\footnote{
	The statement that photons and gravitons are NG bosons is not new and it is discussed in \cite{Ferrari:1971at, Nakanishi:1979fg, Hata:1981nd, Kugo:1985jc}.
} 
Furthermore, it has been shown \cite{Strominger:2014pwa, Pasterski:2015zua, Kehagias:2016zry, Hollands:2016oma, Hamada:2017gdg,Mao:2017wvx} that the conservation of the charges generating the asymptotic symmetries is equivalent to the electromagnetic or gravitational memory effects \cite{ZelPol, Braginsky:1986ia,Braginsky1987,Christodoulou:1991cr, Wiseman:1991ss, Blanchet:1992br, Thorne:1992sdb, Bieri:2013hqa,Tolish:2014bka, Susskind:2015hpa, Bieri:2015jwa,Chu:2016qxp,Tolish:2016ggo,Garfinkle:2017fre,Bieri:2017vni}.

\begin{figure}[htbp]
	\vspace{3mm}
	\begin{center}
		\includegraphics[width=100mm]{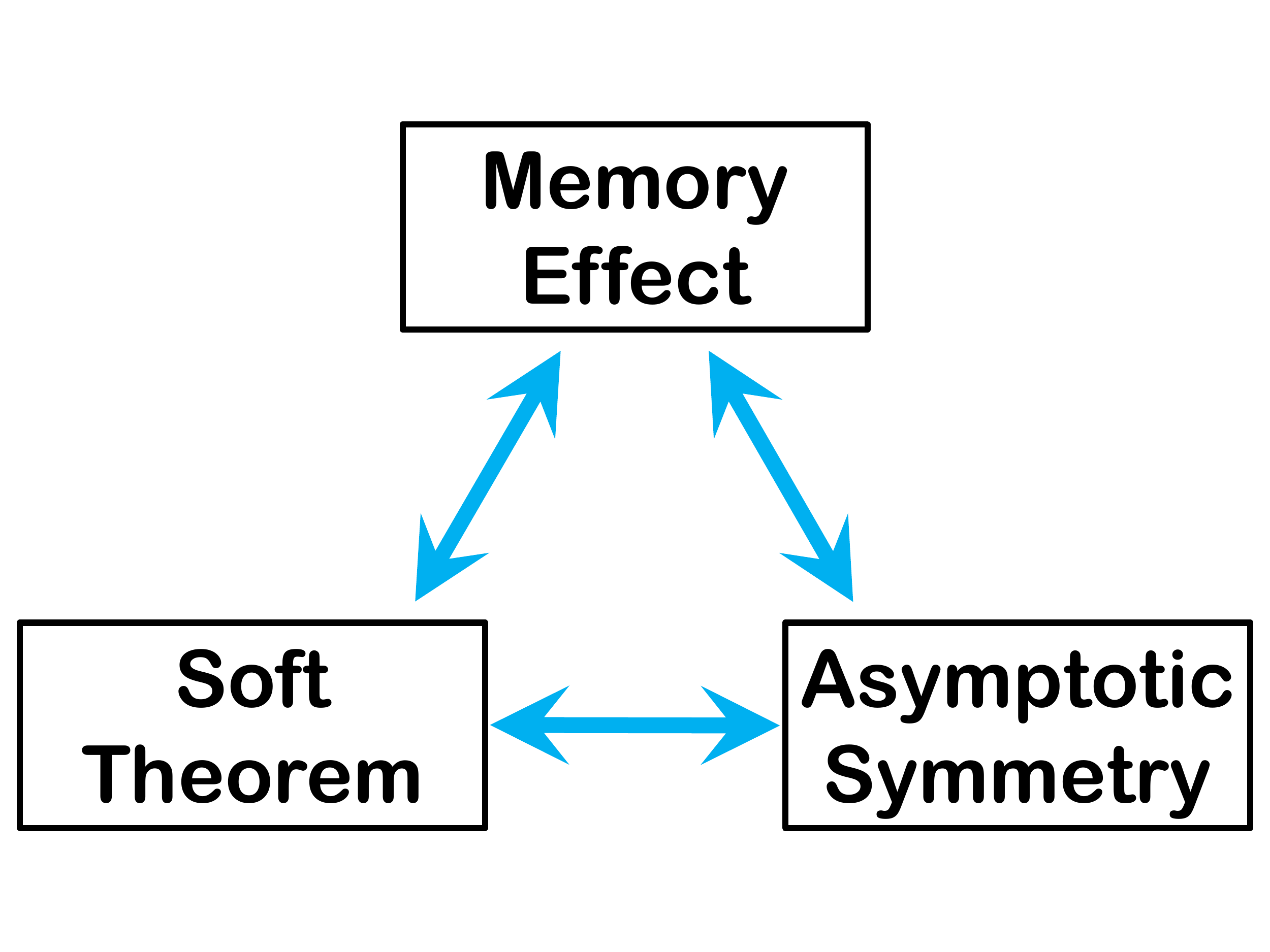}
		\caption{Triangular relation among soft theorem, asymptotic symmetry and memory effect.}
		\label{fig:tri}
	\end{center}    
\end{figure}

Thus, for QED and QG, we have the triangular equivalence relation associated with the infrared dynamics illustrated in figure \ref{fig:tri}.
We expect that such triangular relations hold in other theories with massless particles. 
Actually, the massless scalar theories coupled to massive scalars and fermions are considered in \cite{Campiglia:2017dpg}, and 
it is argued that the soft scalar theorem can be written as the Ward-Takahashi identity and the theories have an infinite number of conserved charges. 
One can show that the charge conservation is equivalent to the scalar memory effect  discussed in \cite{Tolish:2014bka, Garfinkle:2017fre}.   
However, the theories in \cite{Campiglia:2017dpg} suffer from the infrared divergences, and in general, the scalar boson acquires finite mass at loop level. 

In this paper, we consider the case that the massless scalar is a NG boson. Unlike the model considered in \cite{Campiglia:2017dpg}, we consider a theory without infrared divergences, and masslessness of scalar is ensured from the NG theorem. 

The behaviors of scattering amplitudes in the soft limit of a NG boson are different from those in QED and QG since the NG bosons interact only through derivative couplings. Let $\omega$ be the energy of a soft particle. 
In QED and QG, the scattering amplitudes with soft particle, $\omega \to 0$, are factorized into the product of $\mathcal{O}(\omega^{-1})$ soft factor and the amplitudes without soft particle.
For the soft limit of NG boson, the $\mathcal{O}(\omega^{-1})$ factors are absent, and the soft factors start from $\mathcal{O}(1)$. Moreover, even the $\mathcal{O}(1)$ factors often vanish due to so-called Adler's zero \cite{Adler:1964um,Low:2014nga}. 

We consider a specific model that avoids Adler's zero. 
The model contains a complex scalar and a Dirac fermion, and there is a global axial $U(1)$ symmetry, which is spontaneously broken by choosing a vacuum. 
This model may be regarded as a toy model of real pions or axions in the beyond standard model, and therefore we call the associated NG bosons pions. 
In this paper, we first review that the scattering amplitudes with a soft pion give universal $\mathcal{O}(1)$ factors as in \cite{Adler:1966gc}. Then, we rewrite the soft pion theorem as the Ward-Takahashi identity of $S$-matrix by identifying an infinite number of charges which generate an asymptotic symmetry. Furthermore we show that the charge conservation can be interpreted as a pion memory effect, where the information of hard particles is memorized in a shift of $1/r^2$ coefficient of pion fields in future or past null infinity. 
Therefore, the triangular relation in figure \ref{fig:tri} is established for a theory with pions.

The remainder of this paper is organized as follows: 
In section~\ref{sec_model}, we specify a model that we consider in this paper, and the soft pion theorem is presented. Then, the soft theorem is rewritten in a form of the Ward-Takahashi identity, and we find an infinite number of charges generating asymptotic transformations. 
In section~\ref{sec_asympto}, we investigate the asymptotic behaviors of fields near null and timelike infinities, and see the asymptotic transformations. 
In section~\ref{sec_charge_consv}, we argue that the the charge conservation is interpreted as a pion memory effect. We confirm that the memory effect is consistent with the classical asymptotic dynamics.  
Section~\ref{sec_discuss} is devoted to the summary and discussion.
We summarize our conventions in appendix~\ref{app_notation} and coordinate systems in appendix~\ref{app_coord}.  

\section{Soft theorem in $U(1)_A$ model}\label{sec_model}

\subsection{Model}\label{subsec_mode]}
We consider a system of a complex scalar $\Phi$ and a Dirac spinor $\Psi$ interacting as follows:\footnote{See appendix \ref{app_notation} for conventions in our paper. }   
\al{\label{action1}
	\mathcal{L}=
	-\bar{\Psi}{\Slash \p}\Psi
	-\sqrt2 y\paren{\Phi\bar{\Psi}{1+\gamma^5\over2}\Psi +\Phi^*\bar{\Psi}{1-\gamma^5\over2}\Psi}-|\p_\mu\Phi|^2-\frac{\lambda^2}{2}\paren{|\Phi|^2-\paren{v\over\sqrt{2}}^2}^2,
}
where $y$ is the real Yukawa coupling, and $\lambda$ and $v$ are also real couplings.  
This Lagrangian possesses the chiral symmetry, 
\al{\label{chiral_sym}
	&
	\Phi\to e^{i\theta}\Phi,
	\qquad
	\Psi\to e^{-i\theta\gamma^5/2}\Psi.
}
The scalar potential in \eqref{action1} leads to the spontaneous breaking of this $U(1)$ symmetry. 
We choose the vacuum configuration as $\Phi_0= v/\sqrt{2}$, and expand the fields around the vacuum as 
\al{&
	\Phi(x)={1\over\sqrt{2}}\paren{v+\phi(x)}e^{i\uppi(x)/v},
	\qquad
	\Psi(x)=e^{-i\uppi(x)\gamma^5/(2v)}\psi(x), 
}
where $\phi(x)$ and $\uppi(x)$ are real scalar fields, and $\psi(x)$ is the redefined Dirac field.  
The Lagrangian is then given by 
\al{\label{Eq:U(1)A Lagrangian polar}
	\mathcal{L}&=
	-\bar{\psi}{\Slash \p}\psi-y(v+\phi)\bar{\psi}\psi-{1\over2}\br{(\p_\mu\phi)^2+\lambda^2 v^2\phi^2}-{1\over2}\paren{1+{\phi\over v}}^2(\p_\mu \uppi)^2
	\nn&\quad
	+{i\over2v}\paren{\p_\mu \uppi}\bar{\psi}\gamma^\mu\gamma^5\psi
	-\frac{\lambda^2 v}{2} \phi^3-{\lambda^2\over8}\phi^4
	\nn
	&=
	-\bar{\psi}\paren{{\Slash \p}+m}\psi-y\phi\bar{\psi}\psi-{1\over2}\br{(\p_\mu\phi)^2+m_\phi^2\phi^2}-{1\over2}\paren{1+{\lambda \over m_\phi}\phi}^2(\p_\mu \uppi)^2
	\nn&\quad
	+{i y\over 2m}\paren{\p_\mu \uppi}\bar{\psi}\gamma^\mu\gamma^5\psi
	-\frac{m_\phi \lambda}{2} \phi^3-{\lambda^2\over8}\phi^4.
}
Here $m=yv$ and $m_\phi=\lambda v$ are mass of $\psi$ and $\phi$ respectively. $\uppi$ is the NG boson associated with the chiral symmetry \eqref{chiral_sym}, and we call it pion.

\subsection{Soft pion theorem}\label{subsec_soft_theorem}
We now investigate the soft theorem for the NG boson $\uppi(x)$. 
We will see that the soft limit of our NG boson does not lead to the divergence unlike the leading soft theorems for photons \cite{Low:1954kd, GellMann:1954kc}, gravitons \cite{Weinberg:1965nx} and massless scalars \cite{Campiglia:2017dpg}, but it has a universal behavior at the subleading order $\mathcal{O}(1)$.\footnote{In the absence of $\Psi$, $\mathcal{O}(1)$ contributions also vanish due to Adler's zero~\cite{Adler:1964um,Low:2014nga}.} 

First, we summarize the Feynman rules of our model \eqref{Eq:U(1)A Lagrangian polar}. In the interaction picture, fields $\uppi(x), \psi(x), \bar{\psi}(x), \phi(x)$ are expanded as   
\al{\label{eq_a_mode}
	\uppi(x)&=\int {d^3p\over(2\pi)^3}{1\over2 E_{\vb{p}}}\paren{a_{\vb{p}}^{(\uppi)}  e^{i p\cdot x}+a_{\vb{p}}^{(\uppi)\dagger}  e^{-i p\cdot x}},
	\\
	\label{eq_psi_mode}
	\psi(x)&=\int {d^3p\over(2\pi)^3}{1\over2 E_{\vb{p}}}\sum_s\paren{a_{\vb{p}}^s u^s(p) e^{i p\cdot x}+b_{\vb{p}}^{s\dagger} v^s(p) e^{-i p\cdot x}},
	\\
	\label{eq_barpsi_mode}
	\bar{\psi}(x)&=\int {d^3p\over(2\pi)^3}{1\over2 E_{\vb{p}}}\sum_s\paren{b_{\vb{p}}^s \bar{v}^s(p) e^{i p\cdot x}+a_{\vb{p}}^{s\dagger} \bar{u}^s(p) e^{-i p\cdot x}},\\
	\phi(x)&=\int {d^3p\over(2\pi)^3}{1\over2 E_{\vb{p}}}\paren{a_{\vb{p}}^{(\phi)}  e^{i p\cdot x}+a_{\vb{p}}^{(\phi)\dagger}  e^{-i p\cdot x}}.  
}
The annihilation and creation operators satisfy the (anti-)commutation relations:  
\begin{align}
\sqbr{a_{\vb{p}}^{(\uppi)},a_{\vb{p}'}^{(\uppi)\dagger}} &= \sqbr{a_{\vb{p}}^{(\phi)},a_{\vb{p}'}^{(\phi)\dagger}}=2 E_{\vb{p}} (2\pi)^3\delta^3(\vb{p}-\vb{p}')\,, \\
\{a^s_{\vb{p}}, a^{s'\dagger}_{\vb{p}'}\} &= \{b^s_{\vb{p}}, b^{s'\dagger}_{\vb{p}'}\} = 2 E_{\vb{p}} (2\pi)^3 \delta^3(\vb{p}-\vb{p}')\delta^{s,s'}\,.
\end{align}
Then, asymptotic one-particle states are defined by acting creation operators on the free ground state such as $\ket{\vb{p}}_0 =a_{\vb{p}}^{(\uppi)\dagger}\ket{0}_0$.    

From the Lagrangian \eqref{Eq:U(1)A Lagrangian polar}, we obtain the Feynman rules for the perturbative computation of the $S$-matrix elements.  The propagator of each field is  as follows: 
\begin{align}
&\text{propagator of $\uppi$: }
\parbox{25mm}{\includegraphics[width=20mm]{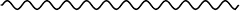}}
=\frac{-i}{p^2-i\epsilon},\\
&\text{propagator of $\psi$: }
\parbox[b]{25mm}{\includegraphics[width=20mm]{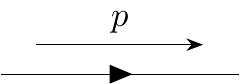}}
=\frac{-\Slash{p}-im}{p^2+m^2-i\epsilon},\\
&\text{propagator of $\phi$: }
\parbox{25mm}{\includegraphics[width=20mm]{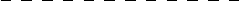}}
=\frac{-i}{p^2+m_{\phi}^2-i\epsilon},
\end{align}
and the interaction vertices including the pions are 
\begin{align}\label{feynman_vertex}
&\parbox{25mm}{\includegraphics[width=20mm]{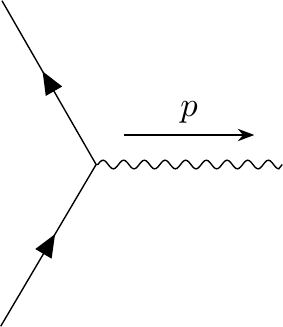}}
=\frac{i y}{2m} \Slash{p}\gamma^5\,, \quad
\parbox{25mm}{\includegraphics[width=20mm]{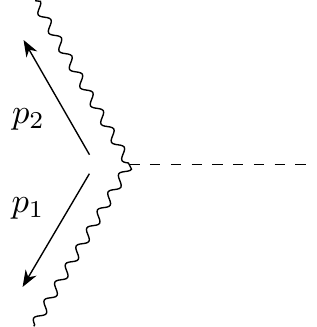}}
=\frac{2i \lambda p_1\cdot p_2 }{m_\phi}\,, \quad
\parbox{25mm}{\includegraphics[width=20mm]{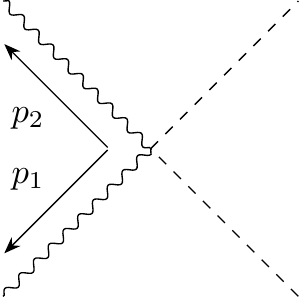}}
=\frac{2i \lambda^2 p_1\cdot p_2 }{m_\phi^2}\,.
\end{align}

\begin{figure}[htbp]
	\vspace{3mm}
	\begin{center}
		\includegraphics[width=50mm]{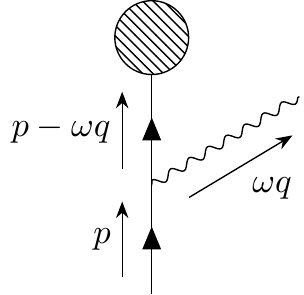}
		\caption{A part of a diagram relevant to the soft limit. The circle represents the other part of the diagram. The momentum $p^\mu-\omega q^\mu$ approaches to on-shell in the soft limit $\omega \to 0$. }
		\label{fig:soft_limit}
	\end{center}    
\end{figure}

We consider scattering processes including an outgoing pion with momentum $\omega q^\mu$ where $q^\mu$ is a normalized null vector $q^\mu=(1,\hat{\vb{q}})$ with $\abs{\hat{\vb{q}}}^2=1$, and take the soft limit $\omega \to 0$. 
Since all of the vertices given in eq.~\eqref{feynman_vertex} are proportional to momenta of the pions, they vanish if the momenta become zero. Thus, the soft limit generally takes the Feynman diagrams to zero unless the limit hits some singularities. 
Such singularities occur only when the external line of the soft pion is attached to the external lines of fermions (or anti-fermions) as Fig.~\ref{fig:soft_limit}.  

The external leg, the vertex and the propagator shown in Fig.~\ref{fig:soft_limit} give a factor 
\begin{align}
\frac{-\Slash{p}+\omega \Slash{q}-im}{(p-\omega q)^2+m^2}\paren{\frac{iy}{2m}\omega\Slash{q}\gamma^5} u^s(p)\,.   
\end{align}
Since $p^\mu$ is the on-shell momentum $(p^2=-m^2)$ and $u^s(p)$ satisfies $\Slash{p}u^s(p)=im u^s(p)$, 
the factor becomes finite in the soft limit $\omega \to 0$ as  
\begin{align}
-\frac{i y}{2m\, p\cdot q}  q^\mu p^\nu \gamma^5\gamma_{\mu\nu} u^s(p)\,,  
\end{align}
where $\gamma_{\mu\nu}= \frac12 [\gamma_\mu, \gamma_\nu]$.
Using the total angular momentum operator of one-fermion which is defined as  
\begin{align}\label{Eq:angular momentum}
J_{\mu\nu} := -i \paren{p_{\mu} \frac{\partial}{\partial p^\nu}-p_{\nu} \frac{\partial}{\partial p^\mu}} -\frac{i}{2} \gamma_{\mu\nu}\,,
\end{align}
and the identity 
\begin{align}
\gamma^5\gamma_{\mu\nu}  = -\frac{i}{2}\epsilon_{\mu\nu\rho\sigma} \gamma^{\rho\sigma}
\qquad (\epsilon_{0123}=1)\,,
\end{align} 
we have
\begin{align}\label{Eq:gamma5 formula}
-\frac{i y}{2m\, p\cdot q} q^\mu p^\nu \gamma^5\gamma_{\mu\nu}
=-\frac{i y}{2m\, p\cdot q} \epsilon_{\mu\nu\rho\sigma} q^\mu p^{\nu} J^{\rho\sigma}\,. 
\end{align}
Notice that only the spin part of eq.~\eqref{Eq:angular momentum} contributes to eq.~\eqref{Eq:gamma5 formula}.
Thus, the soft limit of the diagram is equivalent to considering the diagram without the soft pion and changing the plane-wave spinor $u^s(p)$ into 
\begin{align}\label{eq_tildeu}
u^s(p) \to -\frac{i y}{2m\, p\cdot q} \epsilon_{\mu\nu\rho\sigma} q^\mu p^{\nu} J^{\rho\sigma} u^s(p) \,.
\end{align}

Similarly, the diagram where a soft pion is attached to the external legs of fermions and (anti-fermions) is equivalent to the diagram where $\bar{u}^s(p), v^s(p), \bar{v}^s(p)$ are replaced as follows: 
\begin{align}\label{eq_tildebaru}
\bar{u}^s(p)&\to \frac{i y}{2m\, p\cdot q} \epsilon_{\mu\nu\rho\sigma} \bar{u}^s(p) q^\mu p^{\nu}  J^{\rho\sigma},
\\
\label{eq_tildev}
v^s(p) &\to -\frac{i y}{2m\, p\cdot q} \epsilon_{\mu\nu\rho\sigma} q^\mu p^{\nu} J^{\rho\sigma}  v^s(p)\,,\qquad
\bar{v}^s(p) \to \frac{i y}{2m\, p\cdot q} \epsilon_{\mu\nu\rho\sigma} \bar{v}^s(p)  q^\mu p^{\nu}  J^{\rho\sigma}\,. 
\end{align}
In addition, if the soft pion leg is attached to other than that above, \textit{i.e.}, external legs of $a$ itself and the massive scalar $\phi$, internal lines and vertices, such diagrams vanish in the soft limit, since any internal momenta do not become on-shell as in the proofs of the soft photon and graviton theorems \cite{Weinberg:1965nx}. 
Therefore, the soft pion theorem can be interpreted as acting the operator, 
\begin{align}\label{eq:soft factor}
\frac{i y}{2m\, p\cdot q} \epsilon_{\mu\nu\rho\sigma} q^\mu p^{\nu} J^{\rho\sigma},
\end{align}
to {\it all} external legs including the scalar ones because 
it is trivially zero for scalar particles. 

We express the $S$-matrix element without soft pions as 
\begin{align}
\outst \mathcal{S} \inst,
\end{align} 
where $\outst$ and $\inst$ represent asymptotic multi-particle states with no soft pions, 
and $\mathcal{S}$ denotes the $S$-matrix operator acting on the asymptotic states. 
Then, the $S$-matrix element including an outgoing pion is given by $\outst a_{\omega\hat{\vb{q}}}^{(\uppi)} \, \mathcal{S} \inst$ and satisfies  
\al{\label{softNGthrm}
	\lim_{\omega \to 0} \outst a_{\omega\hat{\vb{q}}}^{(\uppi)} \, \mathcal{S} \inst
	= J^{(1)}(\vb{q}) \outst \mathcal{S} \inst,
}
with 
\begin{align}
J^{(1)}(\vb{q}) = \sum_{k} \frac{-i y \, \eta_k}{2m\, p_k\cdot q} \epsilon_{\mu\nu\rho\sigma} q^\mu p_k^{\nu} J_k^{\rho\sigma},
\end{align}
where $k$ labels hard particles with momentum $p_k^\mu$ and the total momentum $J_k^{\rho\sigma}$, and $\eta_k=1\, (-1)$ for incoming (outgoing) fermions and outgoing (incoming) anti-fermions. 
In the case for the incoming soft pion with momentum $\omega q^\mu$, 
the $S$-matrix element also satisfies the similar equation  
\al{
	\lim_{\omega \to 0} \outst  \mathcal{S} \,  a_{\omega\hat{\vb{q}}}^{(\uppi)\dagger}\inst
	= J^{(1)} (\vb{q}) \outst \mathcal{S} \inst\,. 
}
We thus have the relation\footnote{This relation can be understood from the crossing symmetry too.} 
\begin{align}\label{eq_cross}
\lim_{\omega \to 0} \outst  \mathcal{S} \,  a_{\omega\hat{\vb{q}}}^{(\uppi)\dagger}\inst = \lim_{\omega \to 0} \outst a_{\omega\hat{\vb{q}}}^{(\uppi)} \, \mathcal{S} \inst\,. 
\end{align}

Eq.~\eqref{softNGthrm} is the soft pion theorem that relates the $S$-matrix elements with and without a soft pion. Unlike the soft photon and graviton theorems, the soft limit of the pion does not gives an $\mathcal{O}(\omega^{-1})$ factor. In this sense, the theorem \eqref{softNGthrm} is the subleading soft theorem \cite{Low:1958sn, Gross:1968in, Jackiw:1968zza, Cachazo:2014fwa, Lysov:2014csa}. 
In fact, the right-hand side of \eqref{softNGthrm} is similar to the subleading part of the soft theorem in \cite{Lysov:2014csa}.

\subsection{Soft pion theorem as the Ward-Takahashi identity}\label{sec_soft_ward} 
In this subsection we show that the soft pion theorem \eqref{softNGthrm} can be written as the Ward-Takahashi identity.

We first define the soft charge operator $Q^\text{soft}(\hat{\vb{q}})$ as 
\begin{align}\label{def_soft_charge}
Q^\text{soft}(\hat{\vb{q}}) := -\frac{i}{4\pi}\lim_{\omega \to 0} \paren{a_{\omega\hat{\vb{q}}}^{(\uppi)\dagger}-a_{\omega\hat{\vb{q}}}^{(\uppi)}} \,. 
\end{align}
This Hermitian operator creates (or annihilates) a soft pion, and the coefficients are chosen for later convenience. 
Noting the relation \eqref{eq_cross}, 
we have 
\begin{align}\label{eq_soft1}
\outst \paren{ Q^\text{soft}(\hat{\vb{q}})\, \mathcal{S} - \mathcal{S}\, Q^\text{soft}(\hat{\vb{q}})} \inst = \frac{i}{2\pi} \lim_{\omega \to 0} \outst a_{\omega\hat{\vb{q}}}^{(\uppi)} \, \mathcal{S} \inst.
\end{align}

We next relate the equation \eqref{eq_soft1} to the $S$-matrix element with insertions of a hard charge such as 
\begin{align}\label{eq_hard1}
-\outst \paren{Q^\text{hard}(\hat{\vb{q}})\, \mathcal{S} - \mathcal{S}\, Q^\text{hard}(\hat{\vb{q}})} \inst,
\end{align} 
using the soft theorem \eqref{softNGthrm}. 
If we have such a hard charge operator, we obtain the equation like the Ward-Takahashi identity:  
\begin{align}\label{eq_soft_ward}
\outst \paren{Q(\hat{\vb{q}})\, \mathcal{S} - \mathcal{S}\, Q(\hat{\vb{q}})} \inst =0,
\end{align}  
with 
\begin{align}\label{eq_charge_op}
Q(\hat{\vb{q}}) = Q^\text{soft}(\hat{\vb{q}})+ Q^\text{hard}(\hat{\vb{q}}) \,. 
\end{align}
The hard charge operator $Q^\text{hard}(\hat{\vb{q}})$ actually exists, and is given by 
\begin{align}\label{def_hard_charge}
Q^\text{hard}(\hat{\vb{q}}) := \frac{y}{8\pi m^2}\int\!\! {d^3p\over(2\pi)^3}{1\over2 E_{\vb{p}}}\sum_{s,r} \paren{\bar{u}^r(p) \gamma^5 \frac{q^\mu p^\nu \gamma_{\mu\nu}}{p\cdot q} u^s(p) a^{r\dagger}_{\vb{p}}a^{s}_{\vb{p}}+\bar{v}^s(p) \gamma^5 \frac{q^\mu p^\nu \gamma_{\mu\nu}}{p\cdot q} v^r(p) b^{r\dagger}_{\vb{p}}b^{s}_{\vb{p}}}. 
\end{align}
We now confirm that this hard charge $Q^\text{hard}(\hat{\vb{q}})$ certainly satisfies 
\eqref{eq_soft_ward}. 
Since the soft NG boson insertion is equivalent to changing external legs as \eqref{eq_tildeu}, \eqref{eq_tildebaru} and \eqref{eq_tildev}, taking care of the factor $i/(2\pi)$ in \eqref{eq_soft1} and the sign in \eqref{eq_hard1}, $Q^\text{hard}(\hat{\vb{q}})$ should change external legs of fermions as follows, 
\begin{align}
\label{u_change}
u^s(p)&\to \frac{y}{4 \pi m\, p\cdot q}  q^\mu p^\nu \gamma^5\gamma_{\mu\nu} u^s(p)\,,\qquad
\bar{u}^s(p)\to \frac{y}{4 \pi m\, p\cdot q}  \bar{u}^s(p)  q^\mu p^\nu \gamma^5\gamma_{\mu\nu}\,,\\
\label{v_change}
v^s(p)&\to \frac{-y}{4 \pi m\, p\cdot q} q^\mu p^\nu \gamma^5\gamma_{\mu\nu} v^s(p)\,,\qquad
\bar{v}^s(p)\to \frac{-y}{4 \pi m\, p\cdot q}  \bar{v}^s(p)  q^\mu p^\nu \gamma^5\gamma_{\mu\nu}\,.
\end{align} 
We recall that the Wick contraction between $\psi(x)$ in \eqref{eq_psi_mode} and a  fermionic creation operator $a^{s\dagger}_{\vb{p}}$ gives the plane-wave mode $u^s(p) e^{ipx}$.  
If we apply the hard charge operator \eqref{def_hard_charge} to the incoming one-fermion state $\ket{\vb{p},s}_0=a^{s\dagger}_{\vb{p}} \ket{0}_0$, we have 
\begin{align}
Q^\text{hard}(\hat{\vb{q}})\ket{\vb{p},s}_0 = \frac{y}{8\pi m^2} \sum_r \bar{u}^r(p)  \frac{q^\mu p^\nu \gamma^5\gamma_{\mu\nu}}{p\cdot q} u^s(p) a^{r\dagger}_{\vb{p}} \ket{0}_0\,. 
\end{align}
Using the identity \eqref{spinor_identiy}, 
one can find that 
the Wick contraction between $\psi(x)$ and this state gives the following plane wave 
\begin{align}
\frac{y}{4 \pi m\, p\cdot q}  q^\mu p^\nu \gamma^5\gamma_{\mu\nu} u^s(p) e^{ipx} . 
\end{align} 
One can also find that if the hard charge operator \eqref{def_hard_charge} is acted on one-particle states for incoming anti-fermions and outgoing (anti-)fermions, it correctly  changes the plane-waves as \eqref{u_change} and \eqref{v_change}. 

Therefore, the soft theorem \eqref{softNGthrm} is equivalent to eq.~\eqref{eq_soft_ward} with soft charge \eqref{def_soft_charge} and hard charge \eqref{def_hard_charge}. It implies that the $S$-matrix is invariant under the transformations generated by $Q(\hat{\vb{q}})$, \textit{i.e.}, our theory has the symmetry. 
The charges $Q(\hat{\vb{q}})$ are parametrized by $\hat{\vb{q}}$, and hence we obtain an infinite number of conserved charges. 

Now we obtain the generator of the symmetry which reproduces the soft pion theorem as the Ward-Takahashi identity. In the following sections, we try to investigate the meaning of the symmetry and charge.

Before closing this subsection, we comment that we also have the ``leading" soft pion theorem:  
\begin{align}\label{eq: leading soft}
\lim_{\omega \to 0} \omega\, \outst a_{\omega\hat{\vb{q}}}^{(\uppi)} \, \mathcal{S} \inst
=\lim_{\omega \to 0} \omega\, \outst \mathcal{S}\, a_{\omega\hat{\vb{q}}}^{(\uppi)\dagger}  \inst=0\,, 
\end{align}
which means the absence of $\mathcal{O}(\omega^{-1})$ soft factor. 
It implies that there is another charge $Q^0(\hat{\vb{q}})$ defined as 
\begin{align}\label{def_Q0}
Q^{0}(\hat{\vb{q}}) := -\frac{1}{4\pi} \lim_{\omega\to 0} \omega\, \paren{a_{\omega\hat{\vb{q}}}^{(\uppi)}+a_{\omega\hat{\vb{q}}}^{(\uppi)\dagger}},
\end{align}
which is the same as the soft part of the charge defined in \cite{Campiglia:2017dpg} up to a numerical factor. 
It satisfies 
\begin{align}\label{leading_soft_thrm}
\outst Q^{0}(\hat{\vb{q}}) \, \mathcal{S} \inst = \outst \mathcal{S}\,Q^{0}(\hat{\vb{q}}) \inst=0\,.
\end{align}


\subsection{Ward-Takahashi identity for spontaneously broken symmetry}\label{sec_broken_WT}
In this subsection, we comment that the subleading soft theorem (and the leading soft theorem) can also be derived from the Ward-Takahashi identity for the $U(1)_A$ symmetry \eqref{chiral_sym}. (See, \textit{e.g.}, \cite{Bianchi:2016viy} for relevant discussions.)

The symmetry corresponds to a constant shift of $\uppi$, and the conserved current can be computed  from the Lagrangian \eqref{Eq:U(1)A Lagrangian polar} as 
\begin{align}\label{eq: U(1)_A current}
j_\uppi^\mu:= \frac{\delta \mathcal{L}}{\delta \partial_\mu \uppi} = -\p^\mu \uppi
-\paren{2\frac{\lambda}{m_\phi}\phi+\frac{\lambda^2}{m_\phi^2}\phi^2}\partial^\mu  \uppi
+{i y\over 2m}\bar{\psi}\gamma^\mu\gamma^5\psi.
\end{align}

The Ward-Takahashi identity corresponding to the current conservation $\partial_\mu j_\uppi^\mu=0$ takes the form 
\begin{align}\label{eq: WT U(1)A}
\langle \partial_\mu j_\uppi^\mu (x) \prod_{i=1}^n \uppi(x_i) \prod_{a=1}^A \mathcal{O}(x_a)\rangle =-i \sum_{i=1}^n \delta^4 (x-x_i) \langle \prod_{j\neq i }^n \uppi(x_j) \prod_{a=1}^A \mathcal{O}(x_a)\rangle \,, 
\end{align}
where $\mathcal{O}(x_a)$ denote fields except for pions $\uppi(x)$. 
The soft theorem follows from this identity \cite{Bianchi:2016viy}. 
First, we perform the Fourier transformation from $x^\mu$ to massless on-shell momentum $\omega q^\mu$ and apply the LSZ reduction to the remaining fields. Then, if we take the soft limit $\omega \to 0$, 
the right-hand side of the identity vanishes, and furthermore in the left-hand side, the first term in \eqref{eq: U(1)_A current} gives the amplitude with one soft pion, the second term does not contribute in the soft limit and the last term gives the amplitude with soft factors \eqref{eq:soft factor}. 
Thus, the soft pion theorem \eqref{softNGthrm} is obtained from the Ward-Takahashi identity \eqref{eq: WT U(1)A}. 

The leading soft theorem \eqref{eq: leading soft} is also obtained trivially in the above procedures with multiplying an extra factor $\omega$. 

Therefore, the Ward-Takahashi identities discussed in section \ref{sec_soft_ward} are included in the Ward-Takahashi identity \eqref{eq: WT U(1)A} for the broken $U(1)_A$ symmetry.\footnote{
	The situation is the same as in QED. The leading and subleading soft photon theorem are the consequences of the gauge invariance, and can be derived from the conventional Ward-Takahashi identity of QED. However, by considering the large gauge transformation which depends on the angle, we can obtain the interesting connection with the memory effect. We will comment the advantage of the asymptotic symmetry in Sec.~\ref{sec_discuss}.} 

\section{Asymptotic symmetry}\label{sec_asympto}
In this section, we study the asymptotic transformations generated by the charge operator \eqref{eq_charge_op}. 
As in the large gauge transformations in QED and QG \cite{Strominger:2017zoo}, we consider smeared charges defined as 
\begin{align}\label{soft_charge_ep}
Q^\text{soft} [\epsilon]&:= \int_{S^2} \! d^2 \theta \sqrt{\gamma}\, \epsilon(\theta)\, Q^\text{soft}(\hat{\vb{q}}(\theta))\,,
\\
\label{hard_charge_ep}
Q^\text{hard} [\epsilon]&:=  \int_{S^2} \! d^2 \theta \sqrt{\gamma}\, \epsilon(\theta)\, Q^\text{hard}(\hat{\vb{q}}(\theta))\,, 
\end{align}
where $\epsilon(\theta)$ is an arbitrary function on unit two-sphere $S^2$. 
More precisely, we have two functions $\epsilon^{\pm}(\theta)$ which are respectively defined on spheres in future and past null (or timelike) infinities $\mathscr{I}^\pm$ ($i^\pm$). 
Note that angle coordinates $\theta^A$ in $\mathscr{I}^\pm$ are related to each other antipodally as explained in appendix~\ref{app_coord}\,.\footnote{
Momentum $\hat{\vb{q}}(\theta)$ has the same parametrization for both $\theta^A$. For example, if we take the standard spherical coordinates $(\theta, \varphi)$ for both future and past sphere, the points with the same coordinates $(\theta, \varphi)$ represent the antipodal point of each other. However, $\hat{\vb{q}}$ is parametrized by the same expression $\hat{\vb{q}}=(\sin\theta\cos\varphi, \sin\theta\sin\varphi, \cos\theta)$ for both $(\theta, \varphi)$. 
} 
Thus, $\epsilon^{+}(\theta)$ and $\epsilon^{-}(\theta)$ satisfy the antipodal matching as large gauge parameters in QED \cite{Strominger:2017zoo}. 
In the following, for simplicity, we use $\epsilon(\theta)$ which denotes $\epsilon^{\pm}(\theta)$ collectively.

Similarly, we define the smeared charge of $Q^0$ in \eqref{def_Q0} as 
\begin{align}
Q^{0,\text{out}}[\epsilon]&:= \int_{S^2} \! d^2 \theta \sqrt{\gamma}\, \epsilon(\theta)\, Q^0(\hat{\vb{q}}(\theta))\,,\\
Q^{0,\text{in}}[\epsilon] &:= -\int_{S^2} \! d^2 \theta \sqrt{\gamma}\, \epsilon(\theta)\, Q^0(\hat{\vb{q}}(\theta))\,.
\end{align}

\subsection{Asymptotic behaviors of massless fields}\label{subsec_asympto_massless}
We first consider the asymptotic behaviors of the massless field $\uppi(x)$ at null infinities. The EoM of $\uppi(x)$ is given by 
\begin{align}\label{EOM_a}
\partial^2 \uppi + \partial_\mu j^\mu =0\,,
\end{align}
where 
\begin{align}\label{def_current}
j^\mu:=\paren{2\frac{\lambda}{m_\phi}\phi+\frac{\lambda^2}{m_\phi^2}\phi^2}\partial^\mu  \uppi+\frac{y}{2m} j^{\mu 5}
\qquad \text{with} \quad j^{\mu 5}:= -i\bar{\psi}\gamma^\mu \gamma^5 \psi. 
\end{align}
Note that the EoM \eqref{EOM_a} is nothing less than the conservation of the broken $U(1)_A$ current $j_\uppi^\mu$ defined in \eqref{eq: U(1)_A current}.

Near future null infinity $\mathscr{I}^+$,  
we use the retarded coordinates $(u,r,\theta^A)$ defined in appendix \ref{app_coord}. 
In these coordinates, the solutions of \eqref{EOM_a} decaying at $r\to \infty$ is expanded as\footnote{Since we consider the vacuum corresponding to $\uppi=0$, we do not consider solutions which have $\mathcal{O}(r^0)$ terms.}    
\begin{align}\label{pion_future_null_exp}
\uppi(x) = \frac{\uppi^{(1)}_{\mathscr{I}^+}(u,\theta)}{r} +\frac{\uppi^{(2)}_{\mathscr{I}^+}(u,\theta)}{r^2} +\dots\,.
\end{align} 
Since the source term $\partial_\mu j^\mu$ does not contribute to the leading order, 
the leading term $\uppi^{(1)}_{\mathscr{I}^+}/r$ should be the same as that of 
the free field \eqref{eq_a_mode}. 
Using the stationary phase approximation used in, \textit{e.g.}, \cite{He:2014laa,Campiglia:2017dpg}, 
the free field is expanded as 
\begin{align}
&\int {d^3p\over(2\pi)^3}{1\over2 E_{\vb{p}}}\paren{a_{\vb{p}}^{(\uppi)}  e^{-i E_{\vb{p}} (u+r) +ir\vb{p}\cdot \hat{\vb{x}}(\theta) }+a_{\vb{p}}^{(\uppi)\dagger}  e^{i E_{\vb{p}} (u+r) -ir\vb{p}\cdot \hat{\vb{x}}(\theta) }}
\nn
&=-\frac{i}{8 \pi^2 r} \int^\infty_0\! d \omega \paren{a_{\omega\hat{\vb{x}}(\theta)}^{(\uppi)}e^{-i\omega u}-a_{\omega\hat{\vb{x}}(\theta)}^{(\uppi)\dagger}e^{i\omega u}} +\mathcal{O}(r^{-2})\,,
\end{align}
where $\hat{\vb{x}}=\vb{x}/r$ denotes a point on $S^2$ which is parametrized by $\theta^A$. 
Thus, we obtain 
\begin{align}\label{pi1-future}
\uppi^{(1)}_{\mathscr{I}^+} (u,\theta) =-\frac{i}{8 \pi^2} \int^\infty_0\! d \omega \paren{a_{\omega\hat{\vb{x}}(\theta)}^{(\uppi)}e^{-i\omega u}-a_{\omega\hat{\vb{x}}(\theta)}^{(\uppi)\dagger}e^{i\omega u}}\,.
\end{align}
Similarly, using the advanced coordinates $(v,r,\theta^A)$ near past null infinity $\mathscr{I}^-$, $\uppi(x)$ is expanded as 
\begin{align}
\uppi(x) = \frac{\uppi^{(1)}_{\mathscr{I}^-}(v,\theta)}{r} +\frac{\uppi^{(2)}_{\mathscr{I}^-}(v,\theta)}{r^2} +\dots\,.  
\end{align}
Noting that a point $x^\mu$ in the usual Minkowski coordinates is parametrized as $x^\mu=(v-r,-r\,\hat{\vb{x}}(\theta))$ in the advanced coordinates, one can find that $\uppi^{(1)}_{\mathscr{I}^-}$ is given by 
\begin{align}\label{pi1-past}
\uppi^{(1)}_{\mathscr{I}^-}(v,\theta)=\frac{i}{8 \pi^2} \int^\infty_0\! d \omega \paren{a_{\omega\hat{\vb{x}}(\theta)}^{(\uppi)}e^{-i\omega v}-a_{\omega\hat{\vb{x}}(\theta)}^{(\uppi)\dagger}e^{i\omega v}}\,.
\end{align}

From \eqref{pi1-future} and \eqref{pi1-past}, we find that charges $Q^{0,\text{out}}$ and $Q^{0,\text{in}}$ act on the asymptotic fields as follows: 
\begin{align}
\sqbr{i\, Q^{0,\text{out}} [\epsilon], \uppi^{(1)}_{\mathscr{I}^+} (u,\theta)}= \epsilon(\theta)\,, \qquad
\sqbr{i\, Q^{0,\text{in}} [\epsilon], \uppi^{(1)}_{\mathscr{I}^-} (v,\theta)}= \epsilon(\theta)\,.
\end{align}
Thus, $Q^{0,\text{out}}$ and $Q^{0,\text{in}}$ generate angle-dependent shifts of the asymptotic $1/r$ coefficients of $\uppi(x)$. 
Due to the ``leading" soft theorem \eqref{leading_soft_thrm}, 
we can say that zero modes with respect to $u$ and $v$ of $\uppi^{(1)}_{\mathscr{I}^\pm}$ do not contribute to scattering problems or they are just labels of superselection sectors.

The commutator of soft charge $Q^\text{soft} [\epsilon]$ and $\uppi^{(1)}_{\mathscr{I}^\pm}$ leads to 
\begin{align}
\sqbr{i\, Q^\text{soft} [\epsilon], \uppi^{(1)}_{\mathscr{I}^+} (u,\theta)}=u\, \epsilon(\theta)\,, \qquad
\sqbr{i\, Q^\text{soft} [\epsilon], \uppi^{(1)}_{\mathscr{I}^-} (v,\theta)}=-v\, \epsilon(\theta)\,,
\end{align}
or
\begin{align}
\sqbr{i\, Q^\text{soft} [\epsilon], \partial_u \uppi^{(1)}_{\mathscr{I}^+} (u,\theta)}=\epsilon(\theta)\,, \qquad
\sqbr{i\, Q^\text{soft} [\epsilon], \partial_v \uppi^{(1)}_{\mathscr{I}^-} (v,\theta)}=-\epsilon(\theta)\,.
\end{align}
Therefore, soft charge $Q^\text{soft} [\epsilon]$ generates the angle-dependent  translations of $\partial_u \uppi^{(1)}_{\mathscr{I}^+}$ and $\partial_v \uppi^{(1)}_{\mathscr{I}^-}$. 

\subsection{Asymptotic behaviors of massive fields}\label{subsec_asympto_massive}
To see the asymptotic behaviors of massive fields near timelike infinities $i^\pm$, we use the hyperbolic foliation with coordinates $(\tau, \rho, \theta^A)$ explained in appendix.~\ref{app_coord}. 

Applying the stationary phase approximation to the free forms \eqref{eq_psi_mode}, \eqref{eq_barpsi_mode}, 
Dirac fields $\psi$ and $\bar{\psi}$ are expanded in the far future $\tau\to +\infty$ as 
\begin{align}
\psi(x) = \frac{\psi^{(1)}_+(\rho, \theta)}{\tau^{\frac32}} + \mathcal{O}(\tau^{-\frac52}), \quad \bar\psi(x) = \frac{\bar{\psi}^{(1)}_+(\rho, \theta)}{\tau^{\frac32}} + \mathcal{O}(\tau^{-\frac52}) 
\end{align}
with
\begin{align}
\psi^{(1)}_+(\rho, \theta) &= \frac{\sqrt{m}}{2 (2\pi)^{\frac32}} \sum_s 
\left.\paren{a^s_{\vb{p}} u^s(p) e^{-im\tau-\frac{3 \pi i}{4}} + b^{s\dagger}_{\vb{p}} v^s(p) e^{im\tau+\frac{3 \pi i}{4}}
}\right|_{\vb{p}=m\rho\, \hat{\vb{x}}(\theta)},
\\
\bar\psi^{(1)}_+(\rho, \theta) &= \frac{\sqrt{m}}{2 (2\pi)^{\frac32}} \sum_s 
\left.\paren{a^{s\dagger}_{\vb{p}} \bar{u}^s(p) e^{im\tau+\frac{3 \pi i}{4}} + b^{s}_{\vb{p}} \bar{v}^s(p) e^{-im\tau-\frac{3 \pi i}{4}}
}\right|_{\vb{p}=m\rho\, \hat{\vb{x}}(\theta)}. 
\end{align}
From these expressions, we obtain the following commutators: 
\begin{align}
\sqbr{iQ^\text{hard} (\hat{\vb{q}}), \psi^{(1)}_+(\rho, \theta)} &=\left. \frac{-i y\, q^\mu p^\nu}{4\pi m\, p\cdot q} \right|_{\vb{p}=m\rho\, \hat{\vb{x}}(\theta)} \gamma^5 \gamma_{\mu\nu}\,  \psi^{(1)}_+(\rho, \theta),
\\
\sqbr{iQ^\text{hard} (\hat{\vb{q}}), \bar\psi^{(1)}_+(\rho, \theta)} &=\left. \frac{i y\, q^\mu p^\nu}{4\pi m\, p\cdot q} \right|_{\vb{p}=m\rho\, \hat{\vb{x}}(\theta)} \bar\psi^{(1)}_+(\rho, \theta) \gamma^5 \gamma_{\mu\nu}. 
\end{align}

For the smeared hard charge $Q^\text{hard} [\epsilon]$, the commutators are given by  
\begin{align}
\sqbr{iQ^\text{hard} [\epsilon], \psi^{(1)}_+(\rho, \theta)} &= \frac{-i y}{4\pi m} \Lambda(\rho, \theta;\epsilon) \psi^{(1)}_+(\rho, \theta)\,, 
\\
\sqbr{iQ^\text{hard} [\epsilon], \bar\psi^{(1)}_+(\rho, \theta)} &= \frac{i y}{4\pi m}  \bar\psi^{(1)}_+(\rho, \theta) \Lambda(\rho, \theta;\epsilon)\,, 
\end{align}
where the $\Lambda$ is given by
\begin{align}&
\Lambda(\rho, \theta;\epsilon):= \int_{S^2} \! d^2 \theta' \sqrt{\gamma}\, \epsilon(\theta')\,G(\rho, \theta ;\hat{\vb{q}}(\theta')),
&&
G(\rho, \theta ;\hat{\vb{q}})= \frac{q^\mu Y^\nu \gamma^5 \gamma_{\mu\nu}}{q\cdot Y}.
\end{align}
Here $Y^\mu(\rho, \theta) = ( \sqrt{1+\rho^2},\rho \hat{\vb{x}}(\theta))$, and $G$ satisfies $D^2\, G(\rho, \theta ;\hat{\vb{q}}) =0$, where $D^2$ is Laplacian on $\mathbb{H}^3$  (see appendix~\ref{app_coord}).

Note that when $\epsilon$ is a constant function on $S^2$, we have 
\begin{align}
\Lambda(\rho, \theta;\epsilon) &= -4\pi \, \epsilon \,  \paren{\frac{\sqrt{1+\rho^2}}{\rho}-\frac{\mathrm{Arcsinh}\rho}{\rho^2}} \gamma^5 \gamma_{0i}\, \hat{x}^i(\theta) 
\nn
&=-4\pi \, \epsilon \,  \paren{\frac{\sqrt{1+\rho^2}}{\rho}-\frac{\mathrm{Arcsinh}\rho}{\rho^2}} 
\begin{pmatrix}
\vb*{\sigma}\cdot\hat{\vb{x}}(\theta)&0\\
0&\vb*{\sigma}\cdot\hat{\vb{x}}(\theta)\\
\end{pmatrix}.
\end{align}
which is a spin operator directed to $\hat{\vb{x}}(\theta)$. 
Note that it is known that the factor $\vb*{\sigma}\cdot\hat{\vb{x}}(\theta)$ also appears in the dipole potential of the pion sourced by the fermion, see \textit{e.g.} \cite{Arvanitaki:2014dfa}.

In the same manner, 
Dirac fields $\psi$ and $\bar{\psi}$ are expanded in the far past $\tau\to -\infty$ as 
\begin{align}
\psi(x) = \frac{\psi^{(1)}_-(\rho, \theta)}{(-\tau)^{\frac32}} + \mathcal{O}\paren{(-\tau)^{-\frac52}}, \quad \bar\psi(x) = \frac{\bar{\psi}^{(1)}_-(\rho, \theta)}{(-\tau)^{\frac32}} + \mathcal{O}\paren{(-\tau)^{-\frac52}}, 
\end{align}
with
\begin{align}
\psi^{(1)}_-(\rho, \theta) &= \frac{\sqrt{m}}{2 (2\pi)^{\frac32}} \sum_s 
\left.\paren{a^s_{\vb{p}} u^s(p) e^{-im\tau+\frac{3 \pi i}{4}} + b^{s\dagger}_{\vb{p}} v^s(p) e^{im\tau-\frac{3 \pi i}{4}}
}\right|_{\vb{p}=m\rho\, \hat{\vb{x}}(\theta)},
\\
\bar\psi^{(1)}_-(\rho, \theta) &= \frac{\sqrt{m}}{2 (2\pi)^{\frac32}} \sum_s 
\left.\paren{a^{s\dagger}_{\vb{p}} \bar{u}^s(p) e^{im\tau-\frac{3 \pi i}{4}} + b^{s}_{\vb{p}} \bar{v}^s(p) e^{-im\tau+\frac{3 \pi i}{4}}
}\right|_{\vb{p}=m\rho\, \hat{\vb{x}}(\theta)}. 
\end{align}
Thus, we obtain 
\begin{align}
\sqbr{iQ^\text{hard} [\epsilon], \psi^{(1)}_-(\rho, \theta)} &= \frac{-i y}{4\pi m} \Lambda(\rho, \theta;\epsilon) \psi^{(1)}_-(\rho, \theta)\,, 
\\
\sqbr{iQ^\text{hard} [\epsilon], \bar\psi^{(1)}_-(\rho, \theta)} &= \frac{i y}{4\pi m}  \bar\psi^{(1)}_-(\rho, \theta) \Lambda(\rho, \theta;\epsilon)\,. 
\end{align}
\section{Charge conservation as memory effect}\label{sec_charge_consv}
In this section, 
we give the interpretation of the charge,
\begin{align}\label{charge_epsilon}
Q[\epsilon]=Q^\text{soft} [\epsilon]+Q^\text{hard} [\epsilon]\,,
\end{align}
and argue that the conservation of the charge is equivalent to a pion memory effect. 

\subsection{Expression of charge in terms of the asymptotic fields}
We first express the hard charge $Q^\text{hard} [\epsilon]$ by the asymptotic fields. 
The axial current $j^{\mu 5}$ defined in eq.~\eqref{def_current}  behaves near timelike infinities $i^\pm$ as 
\begin{align}\label{asympt_axial}
j^{\mu 5} &= \frac{j^{\mu 5 (1)}_\pm (\sigma)}{(\pm\tau)^3} + \mathcal{O}\paren{(\pm\tau)^{-4}}\,,\qquad
j^{\mu 5 (1)}_\pm (\sigma):= -i \bar\psi^{(1)}_\pm(\sigma) \gamma^\mu \gamma^5 \psi^{(1)}_\pm(\sigma)\,, 
\end{align}
where $\sigma^\alpha$ represent the coordinates on $\mathbb{H}^3$. 
Using this asymptotic current $j^{\mu 5 (1)}_\pm$, the hard charge $Q^\text{hard}(\hat{\vb{q}})$ is expressed as  
\begin{align}\label{int_Qhard}
Q^\text{hard}(\hat{\vb{q}})=\frac{y}{4 \pi m} \int_{\mathbb{H}^3}\!\!\!d^3 \sigma \sqrt{h}\, Y_\mu :j^{\mu 5 (1)}_\pm: + \frac{y}{4 \pi m}\int_{\mathbb{H}^3}\!\!\!d^3 \sigma \sqrt{h}\, \frac{q_\mu}{Y\cdot q} :j^{\mu 5 (1)}_\pm: \,,
\end{align}
where $h_{\alpha\beta}$  is the metric on $\mathbb{H}^3$ (see appendix~\ref{app_coord}), $Y^\mu(\rho, \theta) = ( \sqrt{1+\rho^2},\rho \hat{\vb{x}}(\theta))$, and $:~:$ denotes  normal ordering. Note that the first term in \eqref{int_Qhard} does not  depend on angle $\hat{\vb{q}}$.

From eq.~\eqref{pi1-future}, creation and annihilation operators of a pion are written as 
\begin{align}
&a_{\omega\hat{\vb{x}}(\theta)}^{(\uppi)}
= 4\pi i \int du e^{i\omega u} \uppi^{(1)}_{\mathscr{I}^+} (u,\theta)\,,
&&a_{\omega\hat{\vb{x}}(\theta)}^{(\uppi)\dagger}
= -4\pi i \int du e^{-i\omega u} \uppi^{(1)}_{\mathscr{I}^+} (u,\theta)\,,  
\end{align}
where we assume that $\omega>0$. 
Thus, the soft charge $Q^\text{soft} [\epsilon]$ in \eqref{soft_charge_ep} can be represented as the integral over future null infinity $\mathscr{I}^+$
\begin{align}\label{soft_charge_future_null}
Q^\text{soft} [\epsilon]=-2 \int_{\mathscr{I}^+} \!\!\! du d^2 \theta \sqrt{\gamma}\, \epsilon(\theta) \uppi^{(1)}_{\mathscr{I}^+} (u,\theta). 
\end{align}
Considering the past null infinity $\mathscr{I}^-$, 
it is also written as 
\begin{align}
Q^\text{soft} [\epsilon]=2 \int_{\mathscr{I}^-} \!\!\! dv d^2 \theta \sqrt{\gamma}\, \epsilon(\theta) \uppi^{(1)}_{\mathscr{I}^-} (v,\theta). 
\end{align} 
Here we insert the asymptotic expansion \eqref{pion_future_null_exp} into EoM \eqref{EOM_a}, and obtain 
\begin{align}
\partial_u \uppi^{(2)}_{\mathscr{I}^+} (u,\theta)= -\frac12 \Delta_{S^2}  \uppi^{(1)}_{\mathscr{I}^+} (u,\theta)\,, 
\end{align}
where the source term $\partial_\mu j^\mu$ is irrelevant at this order. 
Thus, taking the smearing function $\epsilon$ as $\epsilon=\Delta_{S^2}\zeta$ where $\zeta(\theta)$ is an arbitrary function on $S^2$, the soft charge \eqref{soft_charge_future_null} can be written as follows:  
\begin{align}
Q^\text{soft} [\Delta_{S^2}\zeta]&=2 \int_{\mathscr{I}^+} \!\!\! du d^2 \theta \sqrt{\gamma}\, \Delta_{S^2}\zeta(\theta) \uppi^{(1)}_{\mathscr{I}^+} (u,\theta)\nn
&=-4 \int_{\mathscr{I}^+} \!\!\! du d^2 \theta \sqrt{\gamma}\, \zeta(\theta) \partial_u \uppi^{(2)}_{\mathscr{I}^+} (u,\theta)\nn 
&=-4 \int_{S^2} \!\!\! d^2 \theta \sqrt{\gamma}\, \zeta(\theta)\sqbr{ \uppi^{(2)}_{\mathscr{I}^+} (u=\infty,\theta)-\uppi^{(2)}_{\mathscr{I}^+} (u=-\infty,\theta)
}\,.
\end{align} 
It means that the soft charge is related to a shift of  $\uppi^{(2)}_{\mathscr{I}^+}$. 
Since $\uppi^{(2)}_{\mathscr{I}^+}$ is the $1/r^2$ coefficient of pion $\uppi$, we call it dipole-like charge. 
We conclude that the soft charge measures the shift of the dipole-like charge in each angle. 

Therefore, the conservation of charge $Q[\Delta_{S^2}\zeta]$ can be interpreted as the conservation of the following quantity: 
\begin{align}\label{conservation_charge}
-4 \int_{S^2} \!\!\! d^2 \theta \sqrt{\gamma}\, \zeta(\theta)\sqbr{ \uppi^{(2)}_{\mathscr{I}^+} (u=\infty,\theta)-\uppi^{(2)}_{\mathscr{I}^+} (u=-\infty,\theta)
}+\int_{S^2} \!\!\! d^2 \theta \sqrt{\gamma}\, \zeta(\theta) \Delta_{S^2} Q^\text{hard}(\hat{\vb{q}}(\theta)) \,.
\end{align}
This is the memory effect\footnote{
This pion memory effect is small compared with that of QED and QG where the leading effect is $\mathcal{O}(1/r)$. This corresponds to the fact that the soft pion theorem starts from subleading order while the soft photon and graviton theorems starts from the leading order.
} in the sense that the information of the hard charge is memorized in the shift of $\uppi^{(2)}_{\mathscr{I}^+}$. 
In general, $Q^\text{hard}(\hat{\vb{q}}(\theta))$ can be expanded in terms of the spherical harmonics if we require the regularity of $Q^\text{hard}(\hat{\vb{q}}(\theta))$ on $S^2$.
Eq.~\eqref{conservation_charge} implies that $\ell=0$ mode of $\uppi^{(2)}_{\mathscr{I}^+}$ does not change during any scattering process.

\subsection{Classical derivation of the memory effect}
Here we confirm that EoM \eqref{EOM_a}, or the current conservation $\p_\mu j_\uppi^\mu=0$, implies the conservation of \eqref{conservation_charge}.  
We consider the situation that the initial total charge is zero and $\uppi^{(2)}_{\mathscr{I}^+} (u=-\infty,\theta)$ also vanishes. 
Thus, we should have 
\begin{align}\label{conservation_law}
-4 \int_{S^2} \!\!\! d^2 \theta \sqrt{\gamma}\, \zeta(\theta) \uppi^{(2)}_{\mathscr{I}^+} (u=\infty,\theta) +\int_{S^2} \!\!\! d^2 \theta \sqrt{\gamma}\, \zeta(\theta) \Delta_{S^2} Q^\text{hard}(\hat{\vb{q}}(\theta))
=0\,. 
\end{align}

First, we define the axial current in $(\tau, \sigma^\alpha)$ coordinate as
\al{
j^{\tau 5} := \pdv{\tau}{x^\mu} j^{\mu 5}\,,
\qquad
j^{\alpha 5} := \pdv{\sigma^\alpha}{x^\mu} j^{\mu 5}\,, 
}
whose asymptotic behavior in the future infinity $\tau\to +\infty$  is given by
\al{
j^{\tau 5} =: \frac{j^{\tau 5 (1)}_+(\sigma)}{\tau^3} + \mathcal{O}\paren{\tau^{-4}}\,,\qquad
j^{\alpha 5} =: \frac{j^{\alpha 5 (1)}_+(\sigma)}{\tau^4} + \mathcal{O}\paren{\tau^{-5}}\,. 
}
Then, one can show that $\partial_\mu j^{\mu 5}$ behaves in the future infinity $\tau\to +\infty$ as 
\begin{align}\label{source_i+}
\partial_\mu j^{\mu 5} = \frac{1}{\tau^4} D_\alpha j^{\alpha 5 (1)}_+(\sigma) + \mathcal{O}(\tau^{-5})\,,  
\end{align}
where $D_\alpha$ denotes the covariant derivative on the unit hyperbolic space $\mathbb{H}^3$. 
Let pion behave near $\tau\to \infty$ as 
\begin{align}
\uppi= \frac{\uppi^{(1)}_+(\sigma)}{\tau}+\frac{\uppi^{(2)}_+(\sigma)}{\tau^2} + \mathcal{O}(\tau^{-3}). 
\end{align}
Since we have 
\begin{align}
\partial^2 \paren{\frac{\uppi^{(1)}_+(\sigma)}{\tau}}= \frac{1}{\tau^3} (D^2+1) \uppi^{(1)}_+(\sigma)\,,\qquad 
\partial^2 \paren{\frac{\uppi^{(2)}_+(\sigma)}{\tau^2}}= \frac{1}{\tau^4} D^2 \uppi^{(2)}_+(\sigma)\,,
\end{align} 
eq.~\eqref{source_i+} implies that EoM \eqref{EOM_a} leads to  
\begin{align}
\label{eom1}
&(D^2+1) \uppi^{(1)}_+=0,\\
\label{eom2}
&D^2 \uppi^{(2)}_+= -\frac{y}{2m} D_\alpha j^{\alpha 5 (1)}_+\,.  
\end{align}
Note that the first term in $j^\mu$ given by \eqref{def_current} is $\mathcal{O}(\tau^{-7/2})$, and does not contribute in this order. 
Although the source free equation \eqref{eom1} is not important in our analysis, 
one can find its regular solutions, which behave at $\rho\to \infty$ as $\uppi^{(1)}_+ \sim \log\rho/\rho$~\cite{Campiglia:2017dpg}. 

We now solve more important equation \eqref{eom2}. 
It can be solved by Green's function $G_{\mathbb{H}^3}(\sigma,\sigma')$ on $\mathbb{H}^3$ satisfying 
\begin{align}
D^2 G_{\mathbb{H}^3}(\sigma,\sigma') = \frac{1}{\sqrt{h}} \delta^{(3)} (\sigma,\sigma')\,.  
\end{align}
The solution is given by 
\begin{align}\label{def_green_H3}
G_{\mathbb{H}^3}(\sigma,\sigma') = \frac{1}{4\pi} \paren{\frac{Y\cdot Y'}{\sqrt{(Y\cdot Y')^2-1}}+1
}\,,
\end{align}
where $Y^\mu (\sigma)$ is embedding from $\mathbb{H}^3$ to flat space $\mathbb{R}^{1,3}$ given by $Y^\mu(\rho, \theta) = ( \sqrt{1+\rho^2},\rho \hat{\vb{x}}(\theta))$, and $Y^{\prime \mu}= Y^\mu(\sigma')$. 
The last constant in \eqref{def_green_H3} is chosen so that $G_{\mathbb{H}^3}$ decays at $\rho\to \infty$. 
Using this Green's function, $\uppi^{(2)}_+$ is given by\footnote{Although $\uppi^{(2)}_+$ might have the part satisfying the source free equation $D^2 \uppi^{(2)}_+=0$, we ignore it. It may be consistent with the case that we now consider. We assume that $1/r^2$ component of $\uppi$ at spacelike infinity is zero and the change of $\uppi$ arises only from the source term $\partial_\mu j^\mu$.  
}
\begin{align}\label{pi+_solution}
\uppi^{(2)}_+(\sigma') &= -\frac{y}{2 m} \int_{\mathbb{H}^3}\!\!\!d^3 \sigma \sqrt{h}\, 
G_{\mathbb{H}^3}(\sigma',\sigma)\, D_\alpha j^{\alpha 5 (1)}_+(\sigma) \nn
&= \frac{y}{2 m} \int_{\mathbb{H}^3}\!\!\!d^3 \sigma \sqrt{h}\, 
\partial_\alpha G_{\mathbb{H}^3}(\sigma',\sigma)\, j^{\alpha 5 (1)}_+(\sigma)\,,
\end{align}
where we assume that there is no surface term in integration by parts. 

Next, we take the limit $\rho'\to \infty$ in \eqref{pi+_solution}. 
In this limit, we have 
\begin{align}
\lim_{\rho'\to \infty} \partial_\alpha G_{\mathbb{H}^3}(\sigma',\sigma)\, j^{\alpha 5 (1)}_+(\sigma)
=-\frac{1}{4\pi \rho'^2} \paren{\frac{Y_\mu j^{\mu 5 (1)}_+}{(Y\cdot q')^2} +\frac{q'_\mu j^{\mu 5 (1)}_+}{(Y\cdot q')^3} 
}\,, 
\end{align}
where $q^{\prime\mu}$ is a unit null vector parametrized by spherical coordinates $\theta^{\prime A}$ in $\sigma'$ as $q^{\prime\mu}=(1,\hat{\vb{x}}(\theta'))$. 
Furthermore, we have 
\begin{align}
-\frac12 \Delta'_{S^2} \paren{\frac{q^{\prime\mu}}{Y\cdot q'}}=
\frac{Y^\mu}{(Y\cdot q')^2} +\frac{q^{\prime\mu}}{(Y\cdot q')^3}\,, 
\end{align}
where the Laplacian $\Delta'_{S^2}$ acts on $\theta^{\prime A}$. 
Therefore, we obtain 
\begin{align}
\lim_{\rho'\to \infty} \rho^{\prime 2} \,\uppi^{(2)}_+(\rho', \theta')
&= \frac{y}{16 \pi m} \Delta'_{S^2} \int_{\mathbb{H}^3}\!\!\!d^3 \sigma \sqrt{h}\, 
\frac{q'_\mu j^{\mu 5 (1)}_+}{Y\cdot q'}\nn
&= \frac14 \Delta'_{S^2} Q^\text{hard}(\hat{\vb{q}}(\theta'))\,.
\end{align}

The asymptotic fields $\uppi^{(n)}_+$ $(n=1,2)$ near timelike infinity $i^+$ are related to the asymptotic fields $\uppi^{(n)}_{\mathscr{I}^+}$ near null infinity $\mathscr{I}^+$ as 
\begin{align}
\uppi^{(n)}_{\mathscr{I}^+}(u=\infty, \theta)=\lim_{\rho\to \infty} \rho^n\, \uppi^{(n)}_+(\rho,\theta)\,.
\end{align}
Thus, we have 
\begin{align}
\uppi^{(2)}_{\mathscr{I}^+}(u=\infty, \theta)=\frac14 \Delta_{S^2} Q^\text{hard}(\hat{\vb{q}}(\theta))\,. 
\end{align}
This equation is equivalent to the pion memory effect \eqref{conservation_law}. 

\section{Summary and discussions}\label{sec_discuss}
In this paper, we have considered the model where the global axial $U(1)$ symmetry is spontaneously broken.
After the symmetry breaking, there is the NG boson interacting with other particles through the derivative couplings.
Since we may regard this theory as a toy model of pions or axions, we call this NG boson pion. We have investigated the physics of the pions at low energy. It is well-known that the scattering amplitude including the soft pion can be written as the amplitude without soft pion multiplied by a soft factor. 

First, we have reviewed the soft pion theorem, and showed that the subleading soft pion theorem can be written as the simple form, Eq.~\eqref{softNGthrm}. 
Then, we have pointed out that the soft pion theorem can be interpreted as a Ward-Takahashi identity of $S$-matrix under an asymptotic symmetry as Eq.~\eqref{eq_soft_ward}. This suggests that there is a symmetry in the theory which was not known before. The charge of the asymptotic symmetry consists of ``soft" and ``hard" parts, where the soft part creates and annihilates a soft pion, and hard part is responsible for the transformations of hard particles. 
We have clarified the transformation law of each field under the symmetry, and shown that the conserved quantity associated with this transformation is the dipole charge of the pion at every angle. The soft charge represents the difference of the dipole component of the pion before and after the scattering process, and the hard charge corresponds to the dipole flux carried by hard particles. 
The conservation of the charge indicates that the change of the dipole component of the pion is determined by the change of the dipole flux of the hard particle at every angle (see \eqref{conservation_charge}), which is nothing but the pion memory effect in analogy with the electromagnetic and gravitational ones. 
Therefore, there is a deep connection among the soft pion theorem, asymptotic symmetry and the memory effect. In this sense, we have established that the triangular equivalence relation in a NG boson theory, as in the case of QED and QG.

	As we saw in section \ref{sec_broken_WT}, the leading and subleading soft pion theorems can be obtained from the Ward-Takahashi identity for the broken $U(1)_A$ symmetry \cite{Bianchi:2016viy}. In this sense, our asymptotic symmetry is the consequence of the $U(1)_A$ symmetry, although they seem to be different. To identify or investigate the relation between them is an interesting problem. 
	
One may wonder if our asymptotic symmetry is useful because it follows from the usual global $U(1)_A$ symmetry. 
We think that the advantages to consider the asymptotic symmetry are the following two things.
First, the relation to the memory effect is clear. Second, only asymptotic symmetries  might be meaningful if we generalize the theory including quantum gravity. Actually, it is believed (see \textit{e.g.} \cite{Banks:2010zn}) that there is no global symmetry in quantum gravity. Thus, if we consider the theory of pions coupled with quantum gravity, the original $U(1)_A$ is no longer the physical symmetry. However, our asymptotic transformations may still play the role of a symmetry like large gauge transformations of gauge theories.   

Finally, we comment on the possible future directions.
For pions in our world or axions in the beyond the standard model, the axial symmetry is slightly broken by the explicit breaking term and/or the chiral anomaly. 
It would be interesting to study the triangular relation in this context.

In this paper, we have found the asymptotic symmetry of the $S$-matrix. Although we know the transformations of asymptotic fields and the conserved quantity associated to the symmetry, we do not understand this symmetry completely. One way to understand the symmetry more precisely would be considering the canonical quantization of the theory at the constant time or null surface. 

The subleading soft pion theorem is considered in this paper. It is interesting to look for the sub-subleading soft pion theorem, which is not known before to the best of our knowledge. 
The photon and gluon has soft theorem up to the subleading order while the graviton has soft theorem up to sub-subleading order, and the natural question is what happens in the soft pion theorem. 
Furthermore, the soft photon and graviton theorems are completely fixed by the gauge symmetry as well as the Poincare symmetry and locality~\cite{Broedel:2014fsa,Bern:2014vva,Progress}. We expect that a similar argument should hold in the pion case.

Another open issue is the loop correction to the soft theorem.
In the case of gluons and gravitons, if there exist infrared singularities, loop corrections modify the tree level soft theorem~\cite{Bern:2014oka,He:2014bga}. The generalization of their discussions to scalar particles might be interesting. The related discussion was made in Ref.~\cite{Guerrieri:2017ujb}. 

In \cite{Hawking:2016msc}, it is conjectured that black holes have soft hairs corresponding to soft photons and gravitons on the event horizon. 
It is an important problem to examine the possibility that black holes carry soft pion hairs. 

We hope to return to these issues in the future.

\section*{Acknowledgement}
We thank Miguel Campiglia, Hayato Hirai, Yu-tin Huang, Sebastian Mizera and Toshifumi Noumi for stimulating discussions. 
We used TikZ-Feynman \cite{Ellis:2016jkw} for drawing Feynman diagrams. 
This work is supported in part by the Grant-in-Aid for Japan Society for the Promotion of Science (JSPS) Fellows No.16J06151 (YH) and No.16J01004 (SS).

\appendix
\section{Convention and formulas}\label{app_notation}
The metric signature in this paper is $(-,+,+,+)$. 

We use the following representation of the Dirac matrices 
\al{
	\gamma^0=
	\begin{pmatrix}
		-i&0\\
		0&i\\
	\end{pmatrix},
	\qquad
	\gamma^i=
	\begin{pmatrix}
		0&-i\sigma^i\\
		i\sigma^i&0\\
	\end{pmatrix},}
and we have 
\al{&\{\gamma^\mu, \gamma^\nu\} = 2 \eta^{\mu\nu}, 
	\qquad
	\gamma^5= i \gamma^0\gamma^1\gamma^2\gamma^3=
	\begin{pmatrix}
		0&1\\
		1&0\\
	\end{pmatrix},\\
    &
	[\gamma^0, \gamma^i]=
	\begin{pmatrix}
		0 & -2\sigma^i\\
		-2\sigma^i & 0\\
	\end{pmatrix},
	\qquad
	[\gamma^i,\gamma^j]=
	\begin{pmatrix}
		[\sigma^i,\sigma^j]& 0\\
		0&[\sigma^i,\sigma^j]\\
	\end{pmatrix}.
}
The conjugate of the Dirac spinor $\Psi$ is defined as 
\begin{align}
\bar{\Psi}:= i \Psi^\dagger \gamma^0 \,. 
\end{align}
 
The Dirac equations for the positive and negative frequencies in the momentum representation are 
\begin{align}
(i\Slash{p} + m)u^s(p)=0 ,\qquad (-i\Slash{p} + m)v^s(p)=0 .
\end{align}
We normalize the solutions as follows 
\begin{align}
u^s(p)= 
\begin{pmatrix}
\sqrt{E_{\vb{p}} +m} \,\xi^s \\
\frac{p^i \sigma^i}{ \sqrt{E_{\vb{p}} +m}} \, \xi^s\\
\end{pmatrix}\,,
\qquad
v^s(p)= 
\begin{pmatrix}
\frac{p^i \sigma^i}{ \sqrt{E_{\vb{p}} +m}} \, \eta^s\\
\sqrt{E_{\vb{p}} +m} \,\eta^s \\
\end{pmatrix}
\end{align}
with
\begin{align}
\xi^1=\eta^1=\begin{pmatrix}
1\\0\\
\end{pmatrix}\,,
\qquad\xi^2=\eta^2=\begin{pmatrix}
0\\1\\
\end{pmatrix}\,.
\end{align}
They satisfy 
\begin{align}
&\bar{u}^s(p) u^{s'}(p)= 2m \delta^{s s'}\,, 
&&\bar{v}^s(p) v^{s'}(p)= -2m \delta^{s s'}\,,\\
\label{spinor_identiy}
&\sum_{s} u^s_i(p) \bar{u}^s_j(p) =-i(\Slash{p}+ i m)_{ij}\,,
&&\sum_{s} v^s_i(p) \bar{v}^s_j(p) =-i (\Slash{p}-i m)_{ij}\,.
\end{align}

\section{Coordinate systems}\label{app_coord}
In this appendix, we summarize coordinate systems used in this paper. 

For the standard Minkowski coordinates $x^\mu=(t, \vb{x})$, the flat metric is given by 
\begin{align}
ds^2 = -dt^2+\abs{d\vb{x}}^2. 
\end{align}
We also use the spherical coordinates $(t,r,\theta^A)$ where $r=\sqrt{\abs{\vb{x}}^2}$ is the radial distance and $\theta^A$ $(A=1,2)$ are coordinates on unit two-sphere. 
The metric in the coordinates is represented as 
\begin{align}
ds^2 = -dt^2+dr^2+r^2 \gamma_{AB}d\theta^A d\theta^B,  
\end{align}
where $\gamma_{AB}$ is the metric on unit two-sphere. 

It is convenient for studying the behaviors of massless fields near future null infinity $\mathscr{I}^+$ to use the retarded coordinates $(u,r,\theta^A)$, where retarded time $u$ is defined as $u=t-r$.   
The metric in the retarded coordinates takes the form 
\begin{align}
ds^2=-du^2 -2du dr +r^2 \gamma_{AB}d\theta^A d\theta^B. 
\end{align}
Future null infinity $\mathscr{I}^+$ is $r\to \infty$ surface parametrized by $(u,\theta^A)$. 

Similarly, the advanced coordinates $(v,r,\theta^A)$ are useful for working near past null infinity $\mathscr{I}^-$. Advanced time $v$ is given by $v=t+r$. 
As in \cite{Hawking:2016sgy, Strominger:2017zoo}, the angle coordinates $\theta^A$ in the advanced coordinates is not the same as those in the retarded coordinates. We take $\theta^A$ in the advanced coordinates so that they represent the antipodal point of the point $\theta^A$ in the retarded coordinates.\footnote{If we take the standard spherical coordinates $\{x^1=r\sin\theta\cos\varphi, x^2=r\sin\theta\sin\varphi, x^3=r\cos\theta\}$ and set $\theta^A=(\theta,\varphi)$ in the retarded coordinates, $\theta^A$ in the advanced coordinates are given by 
$\theta^A=(\pi-\theta,\varphi+\pi)$. } 
The metric in the advanced coordinates is 
\begin{align}
ds^2=-dv^2 +2dv dr +r^2 \gamma_{AB}d\theta^A d\theta^B. 
\end{align}
Past null infinity $\mathscr{I}^-$ is $r\to \infty$ surface parametrized by $(v,\theta^A)$. 

Near future and past timelike infinities $i^\pm$, we use the hyperbolic foliation of Minkowski space with coordinates $(\tau, \rho, \theta^A)$ considered in \cite{Campiglia:2015qka}. 
For future region $t^2>r^2$ with $t>0$, the coordinates are related to the spherical ones as 
\begin{align}\label{eq_hyper_foliation}
\tau:=\sqrt{t^2-r^2},\qquad \rho:=\frac{r}{\sqrt{t^2-r^2}}. 
\end{align}
The angle coordinates $\theta^A$ is the same as those in the retarded coordinates. 
The metric is given by 
\begin{align}
ds^2=-d \tau^2 + \tau^2\, h_{\alpha\beta}d \sigma^\alpha d \sigma^\beta, 
\end{align}
where $\sigma^\alpha=(\rho, \theta^A)$ are coordinates of unit three-dimensional hyperbolic space $\mathbb{H}^3$ with metric 
\begin{align}
h_{\alpha\beta}d\sigma^\alpha d \sigma^\beta = \frac{d\rho^2}{1+\rho^2} + \rho^2 \gamma_{AB}d\theta^A d\theta^B. 
\end{align}
For past region $t^2>r^2$ with $t<0$, $\tau$ is given by $\tau:=-\sqrt{t^2-r^2}$. 
We take the angle coordinates $\theta^A$ as the same ones in the advanced coordinates. 
The metric is also given by eq.~\eqref{eq_hyper_foliation}.

\bibliographystyle{TitleAndArxiv}
\bibliography{Bibliography}

\end{document}